\title{Methods for a measurement of  $\tau$ polarization  asymmetry in the decay $Z\rightarrow \tau\tau$ at LHC   and determination of the effective weak mixing angle}

\author{
        Vladimir Cherepanov \\
                Institut Pluridisciplinaire Hubert Curien (IPHC)\\
        67037  Strasbourg, France \\
            \and
        Wolfgang Lohmann\\
        III. Physikalisches Institut B, RWTH Aachen University \\
        D-52056 Aachen, Germany \\
}

\date{\today}

\documentclass[12pt]{article}

\usepackage{amsmath}
\usepackage{graphicx}
\usepackage{url}
\usepackage{hyperref}

\begin{document}

\maketitle
\begin{abstract}
In this paper a general method to measure the longitudinal polarization of $\tau$ lepton in the process $pp \rightarrow Z^{0} \rightarrow \tau\tau$ is described. The method of optimal observable allows
to distinguish between the $\tau$ lepton helicity states with maximum possible sensitivity. The optimal observable for the main $\tau$ decay modes ($e/\mu\nu\nu$, $\pi\nu$, $\rho\nu$, $a_{1}\nu$) that can be identified by CMS~\cite{Chatrchyan:2008aa} or ATLAS~\cite{Aad:2008zzm} detectors is described. The analyzing power for $\tau$ polarization can be further gained by taking into account a correlation between longitudinal spin states of $\tau^{+}$ and $\tau^{-}$.   The theoretical systematic uncertainty  on the $\tau$ polarization measured using the $\tau \rightarrow 3\pi\nu$ decays  is estimated to be $(\Delta P_{\tau})_{model} = (1.41 \pm 1.37)\times 10^{-4}$. This value is approximately 50 times smaller than the best uncertainty on $\tau$ polarization measurement given by combination of all LEP results. An approach to propagate the measured $\tau$ polarization averaged over the $Z^{0}$ lineshape and parton density functions to the electroweak parameters is also discussed. 

\end{abstract}
\section{Introduction}
\label{Introduction}
The weak mixing angle $\sin\theta^{2}_{eff}$ has been measured at different experiments. The two most precise measurements are
obtained by LEP and SLD experiments~\cite{ALEPH:2005ab} . The two results show a disagreement of about three standard deviations. The measurement of $\sin\theta^{2}_{eff}$
has also been performed at CDF and D0 experiments at Tevatron~\cite{Aaltonen:2018dxj} which yields so far the most
precise measurement at hadron collider.  At LHC the measurement of the weak mixing angle has been reported by CMS, ATLAS and LHCb 
experiments~\cite{Chatrchyan:2011ya,Aad:2015uau,Aaij:2015lka} where $\sin\theta^{2}_{eff}$ is extracted 
from the forward-backward charge assymetry ($A_{FB}$) of the process $pp\rightarrow Z/\gamma*\rightarrow ll$ ($l=e,\mu$). The LEP precision of these measurements is not yet
achieved. An independent and complementary measurement at LHC can be performed extracting the weak mixing angle from the polarization of $\tau$ leptons produced
in the process $pp\rightarrow Z/\gamma*\rightarrow \tau\tau$. An additional interest in measuring $\tau$ polarization is the determination of the ratio 
of vector to axial-vector neutral couplings for $\tau$ leptons, which allows to test the lepton universality of the weak neutral current. 
With currently reached experimental precision these couplings are the same for all leptons

The difference of the  neutral weak  couplings to the  right- and left-handed fermions results 
in the polarization of fermion-antifermion pairs produced in the decay of the Z boson.
The  $\tau$ polarization is defined as $P_{\tau} = \frac{\sigma_{+} - \sigma_{-}}{\sigma_{+} + \sigma_{-}}$, where $\sigma_{+}$ and $\sigma_{-}$
are the  cross sections of the production of the  $\tau^{-}$ with positive and negative helicity, respectively.
The differential  cross section of the process $q\bar{q} \rightarrow Z \rightarrow \tau^{+}\tau^{-}$ in the lowest order can be expressed as~\cite{Eberhard:1989ve}:
\begin{equation}\label{crossection}
\begin{split}
\frac{d\sigma}{d\cos\theta_{\tau}} = F_{0}(s)(1+\cos^{2}\theta_{\tau}) + 2F_{1}(s)\cos\theta_{\tau} -\\
 h_{\tau}[F_{2}(1+\cos^{2}\theta_{\tau}) + 2F_{3}\cos\theta_{\tau}].
\end{split}
\end{equation}
Here $\theta_{\tau}$ is the scattering angle of the $\tau^{-}$ with respect to the anti-quark momentum in the rest frame of the 
Z boson, $h_{\tau}$ the helicity of the $\tau^{-}$ and
$\sqrt{s}$ the center-of-mass energy of the quark anti-quark pair.  The $F_{i}(s)$ are structure functions depending on the neutral current vector and axial-vector couplings of the initial quark flavour and the $\tau$ lepton.
The total cross section is: \\
\begin{equation}\label{eq12}
\sigma  = \Sigma_{h_{\tau}}\int\frac{d\sigma}{d\cos_{\theta_{\tau}}}d\cos\theta_{\tau}. 
\end{equation}

From the cross section~(\ref{crossection}) the $\tau$ polarization asymmetry, $P_{\tau}$, as a function of $\sqrt{s}$ is given as:

\begin{equation} \label{consts}
 P_{\tau} = \frac{1}{\sigma}[\sigma(h_{\tau} = +1) - \sigma(h_{\tau} = -1)] = -\frac{F_{2}(s)}{F_{0}(s)}, 
\end{equation}
The cross sections include contributions from 
$Z^{0}$ exchange, photon exchange and photon-$Z^{0}$ interference.
The contributions from the photon exchange cancel in the asymmetries. On top of the $Z^{0}$ resonance ($\sqrt{s} = M_{Z^{0}}$), the following holds:
\begin{equation} \label{consts1}
P_{\tau} =  -A_{\tau}
\end{equation}
where $A_{\tau} = \frac{2v_{\tau}a_{\tau}}{v^{2}_{\tau} + a^{2}_{\tau}}$ and $v_{\tau}$ and $a_{\tau}$ the effective 
neutral current vector and axial vector couplings of the $\tau$ lepton. At  tree-level the couplings are:

\begin{equation}\label{weakcouplings}
  \begin{split}
    v_{\tau} &= I^{3}_{W} - 2Q\sin^{2}\theta_{W},\\
    a_{\tau} &= I^{3}_{W},\\
  \end{split}
\end{equation}
where Q is the electrical charge, $I^{3}_{W}$ the third component of the weak isospin of the left-handed $\tau$  lepton and $\sin^{2}\theta_{W}$  is 
the weak mixing angle. Taking into account  of the smallness of $v_{\tau}$ comparing to $a_{\tau}$ the relation between
$\tau$  lepton polarization, the ratio of weak couplings of $\tau$ lepton and the effective weak mixing angle is:


\begin{equation}\label{eqnumber5}
P_{\tau} \approx 2\frac{v_{\tau}}{a_{\tau}} = 2 - 8\sin^{2}\theta_{W},
\end{equation}

A correct propagation of the measured $\tau$  lepton polarization to the ratio of the weak $\tau$ couplings taking into account the flavour of the initial quark, dependency of
the $\tau$ polarization from the center-of-mass energy of the initial quark- anti-quark pair  and effect of quantum corrections  is discussed  in Sec.~\ref{determineSinEff}.

In this article a general approach to measure $\tau$ polarization with a maximum possible sensitivity in main $\tau$ lepton decay channels ($\tau \rightarrow \pi\nu, \rho\nu, a_{1}\nu$) that can identified by CMS and ATLAS detector is described.

\section{Reconstruction of $\tau$ - pair kinematic at LHC}\label{kinreco}
The knowledge of $\tau$  lepton kinematic is essentially important for the spin analysis. An angle between the $\tau$  lepton and a neutrino 
 in its decay is a powerful spin analyzer. This angle can be reconstructed unambiguously 
only if both the total momentum and the direction of the $\tau$ lepton are reconstructed. Analysis of the longitudinal $\tau$ spin in assumption that the full kinematic
is available is discussed in Sec.~\ref{PolarVectorAndOO}.  Unlike in $e^{+}e^{-}$ case in proton-proton collisions there is no beam energy constraint, however in principle it is 
possible to place a reasonable estimate on the full kinematic of the $\tau$ pair in the process $pp \rightarrow Z \rightarrow \tau\tau$, e.g. the momentum escaped with neutrinos.

\begin{itemize} 

\item The Missing Mass Calculator (MMC)~\cite{Elagin:2010aw}  method assumes that there is no other neutrinos in the event other than the ones from decays of both $\tau$ leptons. Within this
assumption the invisible momenta caried by neutrinos can be estimated from the measured missing transverse energy.  However, in this approach the number of unknowns (from 6 to 8, depending on
the number of neutrinos in decays of both $\tau$ leptons) exceeds the number of constraints. An additional information that is used to find an unambiguous solution for neutrino momentum is an
angular distance  between the neutrino(s) and the visible $\tau$ decay products. This additional constraints for neutrinos from the decays  of both $\tau$ leptons  are incorporated as a probability density functions in an event likelihood.  This probability is calculated for every point in the phase space allowed by the missing transverse energy constraint and the most probable value is taken as a momenta estimate.

\item {\it SVFit}~\cite{Bianchini:2014vza} algorithm has been developed for CMS  $H \rightarrow \tau\tau$  searches~\cite{PhysRevLett.106.231801}. The algorithm computes for each event a likelihood function $P(M_{\tau\tau})$ which quantifies the level of compatibility of a mother particle mass hypothesis $M_{\tau\tau}$ with measured momenta of the visible $\tau$ decay products plus the missing transverse energy reconstructed in the event. $M_{\tau\tau}$ values  are  reconstructed  by  combining  the  measured  observables for missing transverse energy  with  a probability model, which  includes terms for $\tau$ decay kinematics, for the missing transverse energy resolution and with a constraint that visible $\tau$ decay products are equal to the observed.

\item The Global Event Fit~\cite{Cherepanov:2018npf}  algorithm has been developed for the decays $Z \rightarrow \tau\tau  \rightarrow  X + a_{1}\nu$, with $a_1$ resonance decaying into three charged pions.
The high multiplicity of proton-proton collisions allow to reconstruct the point of interaction, i.e. the point of $\tau$ pair production. The point of $\tau$ lepton that decays 
to $a_{1}$  is reconstructed using three charged tracks from $a_{1}$ decay, requiring all three tracks to originate from the common space point. 
 These two points provide a robust way to reconstruct the flight direction of the first $\tau$. 
The further method consist in iterative minimization of the likelihood built from the angular constraints on the flight direction of the second $\tau$ derived from the  event decay topology and an  explicit assumption on the parent particle of a $\tau$ pair, $M_{Z}$ or $M_{H}$.

The described methods demonstrated a good performance in the estimation of the invisible neutrinos momenta and can be also used for measurement of $\tau$ polarization.

\end{itemize}

\section{Angular analysis}
\label{angularanalysis}

The spin of $\tau$ is transformed into the total angular momentum of its decay products and therefore it reveals through the angular distributions of
the decay products relative to each other and to the flight direction of the $\tau$ lepton.

\begin{figure}[ht]
  \begin{center}
    \includegraphics[width=0.5\textwidth]{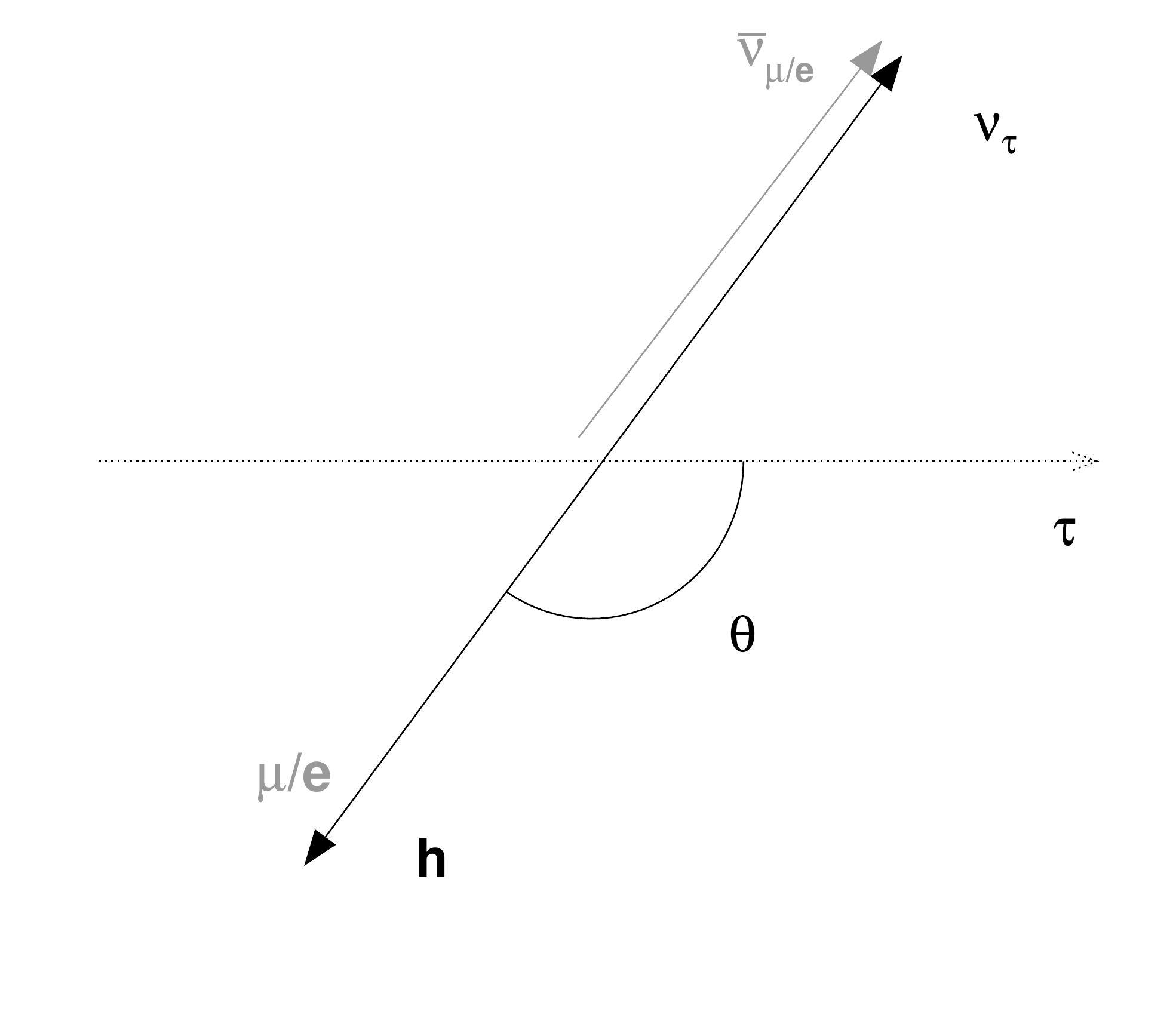}
    \caption{ Definition of angle $\theta$ in the decays $\tau \rightarrow h\nu \ \ (h = \pi,\ \ \rho,\ \ a_{1})$ and $\tau \rightarrow e/\mu \nu\bar{\nu}$. For the leptonic decay the picture shows the high energy lepton production when both neutrinos are emitted in one direction. }
    \label{thetaangle}
  \end{center}
\end{figure}

 The first angle to consider is the angle $\theta$ between the direction
of flight of the $\tau$ in the laboratory frame  and the direction of the hadron, $h$ or direction of the lepton in case of the  leptonic $\tau$ decay as it is seen from the $\tau$ rest frame:

\begin{equation}\label{tyhetaast}
\cos\theta = {\bf n_{\tau}}\cdot{\bf p},
\end{equation}
where ${\bf n_{\tau}}$ is the vector pointing along the $\tau$ direction and ${\bf p}$ is a unit vector pointing along the  momentum of the hadrons or the lepton in the $\tau$ rest frame. The angle $\theta$ is  schematically shown in Fig. \ref{thetaangle}.

In leptonic $\tau$ decays it is possible to reconstruct only the sum of neutrinos momentum and therefore the angle $\theta$ is the only spin sensitive quantity that can be experimentally accessed in CMS or ATLAS detectors.
In case of pseudoscalar $\tau \rightarrow \pi\nu$ decay the angle $\theta$ carries all spin information. 

The angle $\theta$ in production of spin-one resonance has some common features and further is discussed together, denoting $V$ to be  the vector  $\rho \ \ (J^{P} = 1^{-})$ or axial-vector $a_{1}  \ \ (J^{P} = 1^{+})$ resonance.

The $\tau \rightarrow V\nu$ decay channel offers the kinematic simplicity of a two-body decay, like the $\tau^{-} \rightarrow \pi^{-}\nu$, but with with
more complicated dynamic since the resonance $V$ can have longitudinal and transverse spin  states. Conservation of angular momentum allows the $V$ resonance to have $\lambda_{V}$  =0 
or -1.  If the $\tau$ lepton is in the right-handed state the $V$ resonance tends to be in longitudinally polarized state ($\lambda_{V}$ =0) and oppositely if $\tau$ is left-handed
the $V$ is preferably is transversely polarized ($\lambda_{V}$ = -1). Combining the spin amplitudes  for all possible configuration of $V$ resonance and $\tau$ helicities, one gets:

\begin{equation} \label{combinedampl}
\frac{1}{\Gamma}\frac{d\Gamma}{d\cos\theta} \propto 1 + \alpha_{V}h_{\tau}\cos\theta,
\end{equation}
where  the dilution factor  $\alpha_{V} = \frac{|M_{L}|^{2} - |M_{T}|^{2}}{|M_{T}|^{2} + |M_{L}|^{2}} = \frac{m^{2}_{\tau} - 2m^{2}_{V}}{m^{2}_{\tau} + 2m^{2}_{V}} $  is a 
result of the presence of the transverse $V$ resonance amplitude. The value of the factor $\alpha$ characterizes the sensitivity of the $\cos\theta$ observable. Fore comparison, in the $\tau$ decay to  $a_{1}$, $\alpha_{a_{1}} = 0.021$, for  $\rho$, $\alpha_{\rho}= 0.46$ and in the pion decay $\alpha_{\pi} = 1$.   Consequently, the sensitivity to the $\tau$ helicity  in the decay $\tau \rightarrow V \nu$ is strongly reduced if only the  $\cos\theta$ angle is analyzed.  The loss of sensitivity due to factor $\alpha_{V}$ can be compensated if it is distinguished whether  the $V$ resonance is in a transverse or longitudinal state.  

The spin of the $V$  is transformed into the total  angular momentum of the decay products and thus can be retrieved  by analyzing the subsequent  $V$ decay.

\begin{figure}[ht]
  \begin{center}
    \includegraphics[width=0.5\textwidth]{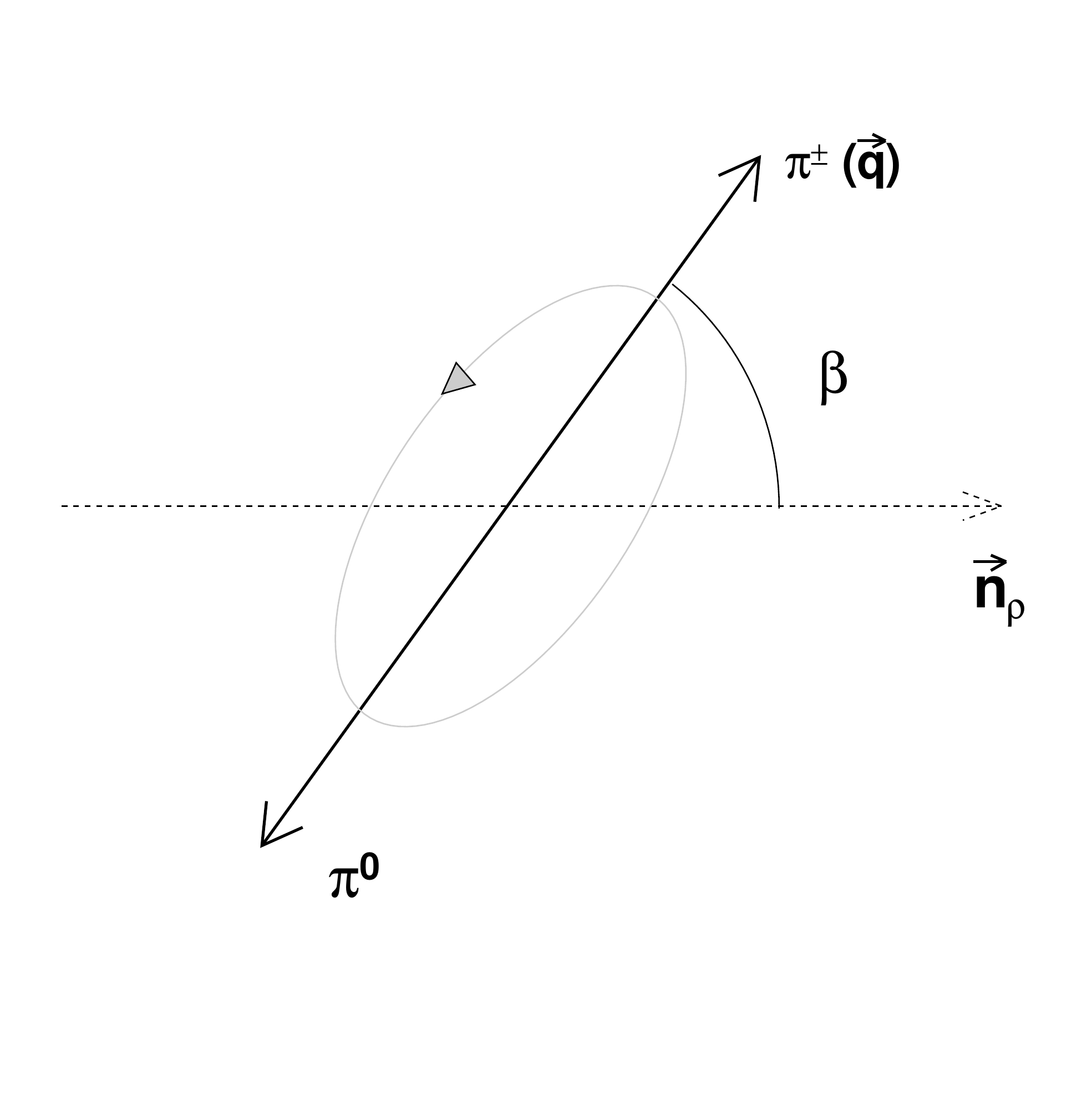}
    \caption{Definitions of angle $\beta$ in the decay $\tau \rightarrow \rho\nu$. }
    \label{rhobetaplot}
  \end{center}
\end{figure}

There are two additional angles in the decay $\tau \rightarrow \rho\nu \  \ (\rho \rightarrow \pi\pi)$.  $\beta$ denotes the angle between the direction of charged pion and the direction of
$\rho$ as seen in the $\rho$ rest frame:

\begin{equation} \label{betaequ}
\cos\beta = {\bf q} \cdot {\bf n_L},
\end{equation}
where ${\bf q}$ is a unit vector along direction of charged pion in the $\rho$ rest frame, ${\bf n_L}$ direction of the laboratory frame. 
The angle $\beta$ in $\rho$ decay is shown in Fig.~\ref{rhobetaplot}. In case of $\lambda_{\rho} = 0$  the angle $\beta$ tends to small values and to large values if $\lambda_{\rho} = -1$.

\begin{figure}[ht]
  \begin{center}
   \hspace{5cm} \includegraphics[width=0.80\textwidth]{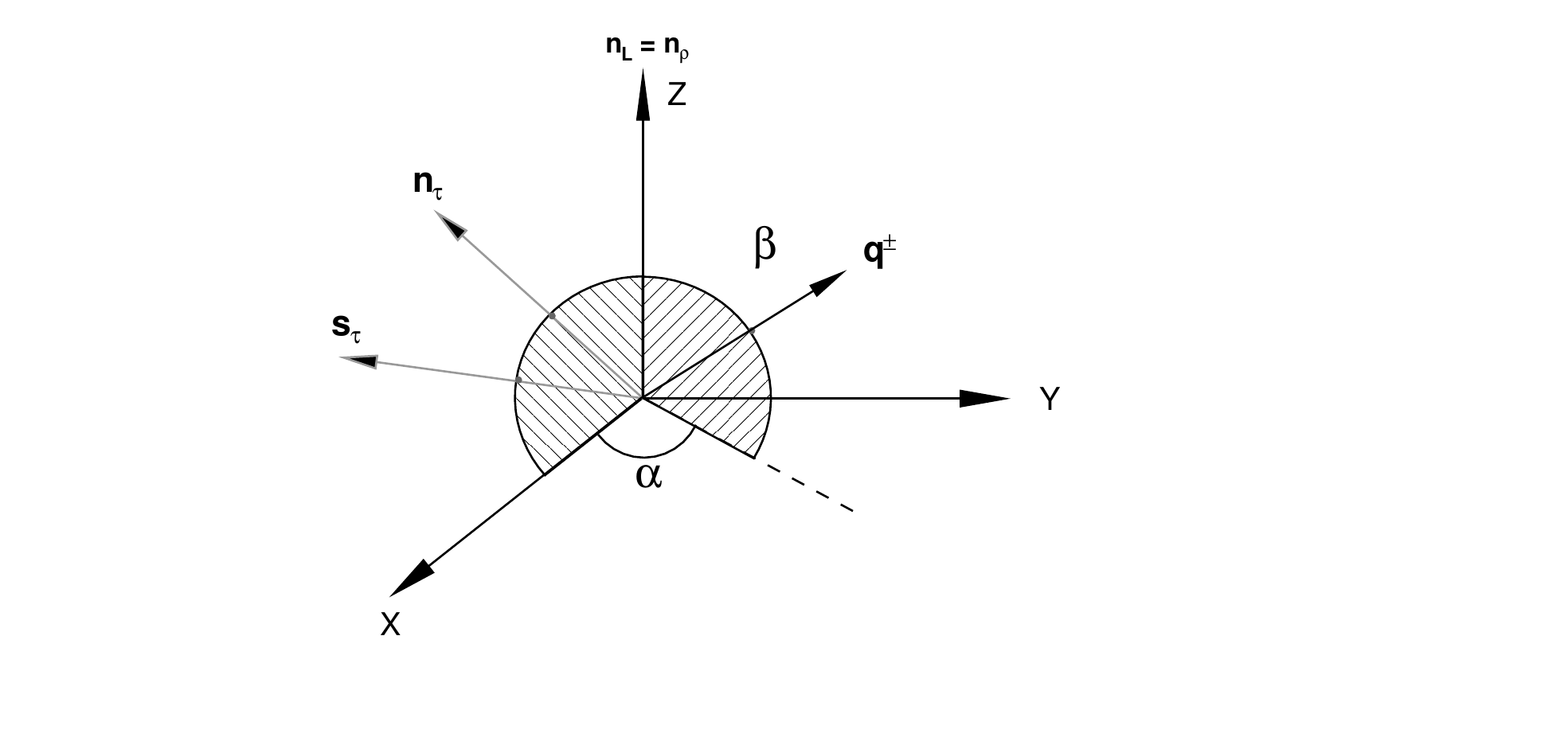}
    \caption{Definition of angle $\alpha$ in the decay $\tau \rightarrow \rho\nu$.   ${\bf n_{\tau}}$ is the direction of $\tau$ lepton, ${\bf s_{\tau}}$ the spin of $\tau$,   ${\bf n_{L}}$  direction of the laboratory frame and   ${\bf q}$  the direction of the charged pion in the $\rho$ rest frame. }
    \label{alpharho}
  \end{center}
\end{figure}

The second angle  $\alpha$ is defined by the angle between two planes spanned by vectors (${\bf n_{L}}, \ \  {\bf n_{\tau}}$) and (${\bf n_{L}}, \ \  {\bf q}$) respectively:

\begin{equation} \label{alpharhoeq}
\cos\alpha = \frac{({\bf n_{L}}\times{\bf n_{\tau}})\cdot ({\bf n_{L}}\times{\bf q}) } {|{\bf n_{L}}\times{\bf n_{\tau}}| \cdot |{\bf n_{L}}\times{\bf q}|},
\end{equation}
where ${\bf n_{\tau}}$ is the direction of $\tau$ lepton,    ${\bf n_{L}}$  direction of the laboratory frame and   ${\bf q}$  the direction foo the charged pion. All vectors are defined in the $\rho$ rest frame. 
 The angle $\alpha$ contains a correlation between the spin of $\tau$ lepton and decay products of the $\rho$ meson. The angle is shown in Fig.~\ref{alpharho}.

\begin{figure}[ht]
  \begin{center}
    \includegraphics[width=0.7\textwidth]{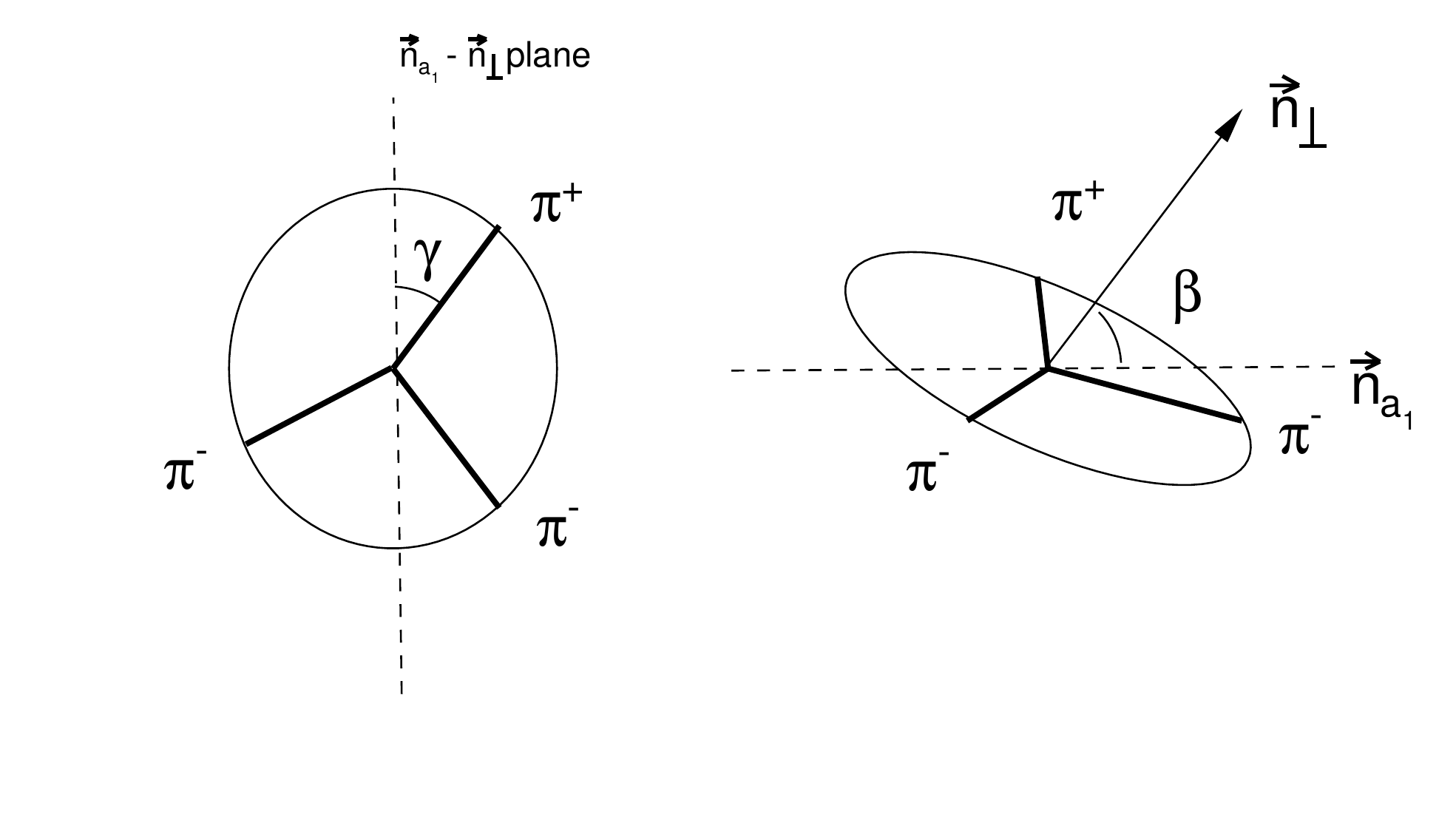}
    \caption{Definitions of angles $\beta$ (\textit{right}) and $\gamma$ (\textit{left}) in the $a_{1}$ rest frame. $\vec{n}_{\perp}$  defines the 3$\pi$ plane in the $a_{1}$ rest frame, and $\vec{n}_{a_{1}}$ the direction of the laboratory frame. }
    \label{betagammascetch}
  \end{center}
\end{figure}

In the decay $\tau^{-}~\rightarrow~a_{1}\nu$  there are three angles in addition to angle $\theta$.  All three angles are described in the 
$a_{1}$ rest frame.  An angle $\beta$ is defined  as the angle  between the normal to the $3\pi$ decay plane and the $a_{1}$ flight direction. 
This is analogous to angle $\beta$ in the $\tau \rightarrow \rho\nu$ decay channel. The angle $\gamma$  describes the relative orientation of the pions in their decay plane. Both angles are shown in Fig.~\ref{betagammascetch} for an illustration. Quantitatively these angles can be defined as: 

\begin{equation} \label{betagammaa1restframe}
\begin{split}
\cos\beta &= {\bf n_{\perp}}\cdot{\bf n_{a_{1}}}, \\
\sin\gamma &= \frac{ ({\bf n_{\perp}}\times{\bf n_{a_{1}}}){\bf q_{3}}}{|{\bf n_{\perp}}\times{\bf n_{a_{1}}}|}, \\
\end{split}
\end{equation}
 where ${\bf n_{a_{1}}}$ denotes direction of the $a_{1}$ in the laboratory frame, $ {\bf n_{\perp}}$ the normal to the $3\pi$ plane in the $a_{1}$ rest frame,  and ${\bf q_{3}}$  the unit vector along the direction of the $\pi^{+}$ ($\pi^{-}$) if the final state is $\pi^{-}\pi^{-}\pi^{+}$ ($\pi^{+}\pi^{+}\pi^{-}$). The last angle $\alpha$ is also similar to the same angle in the $\rho$ decay and can be calculated from Eq.~(\ref{alpharhoeq}) substituting ${\bf q}  \rightarrow {\bf n_{\perp}}$.  The knowledge of all these angles in decays $\tau \rightarrow \rho\nu$, $\tau  \rightarrow a_{1}\nu$ provides the full description of these decays and allows the maximum sensitivity in the spin analysis of the $\tau$ lepton.

In general, there is no fundamental problem in extracting the information on $\tau$ helicity state from a multidimensional angular distribution, however, often it is not practical. 
The more practical approach is to define one-dimensional quantity which is a combination of the spin sensitive angles described above and carries the full  analyzing power. The method is described in the next section.

\section{Polarimetric vector and optimal observable}\label{PolarVectorAndOO}

In general case the differential widths of any decay of a polarized $\tau$ lepton has a form~\cite{Tsai:1971vv,Jadach:1993hs}:

\begin{equation}\label{decaydistr12}
d\Gamma = \frac{|\bar{M}|^{2}}{2m_{\tau}}(1-h_{\mu}s^{\mu})dLips,
\end{equation}
where $s$ is a four-vector of the spin of $\tau$, $h$ is the polarimetric vector.
As it was discussed in  Sec.~\ref{kinreco}  practically it is possible to reconstruct the rest frame of the $\tau$ lepton, in this frame $s_{0}$ =0 (the choice for $h_{0}$ is therefore arbitrary) and then:

\begin{equation}\label{decaydistr2}
\frac{d\Gamma}{d\cos\theta_{h}} \propto \frac{1}{2}(1 + P_{\tau}\cos\theta_{h}),
\end{equation}
where $P_{\tau}$ is the $\tau$ polarization or a helicity state for a given $\tau$ decay, $\cos\theta_{h}$ is an angle between polarimetric vector $\vec{h}$
and the direction of $\tau$ lepton $\vec{n}_{\tau}$ as seen from the $\tau$ rest frame.

In the semileptonic $\tau$ decay $\tau(P) \rightarrow X(Q) + \nu_{\tau}(N)$, with $P$, $Q$ and $N$ being a four-momentum of $\tau$ lepton, $X$ and neutrino,  a general expression for a polarimetric vector in the Standard Model (massless neutrinos, $v^{2} = a^{2}$ and $\gamma_{va}=-1$) is given by~\cite{Jadach:1993hs}:

\begin{equation}\label{polarimetricvector}
h_{\mu} = \frac{H_{\mu}}{b},
\end{equation}
where

\begin{equation}\label{polarimetricH}
H_{\mu} = \frac{1}{M}(M^{2}\delta^{\nu}_{\mu} - P_{\mu}P^{\nu})(\Pi^{5}_{\nu} - \gamma_{va}\Pi_{\nu}),
\end{equation}

\begin{equation}\label{polarimetricb}
b = P^{\mu}(\Pi_{\mu} - \gamma_{va}\Pi^{5}_{\mu}),
\end{equation}

\begin{equation}\label{polarimetricPIVEC}
  \Pi_{\mu} = 2\Big((J^{\ast}N)J_{\mu} + (JN)J_{\mu}^{\ast} - (J^{\ast}J)N_{\mu}\Big),
\end{equation}

\begin{equation}\label{polarimetricPIAX}
\Pi^{5\mu} = 2Im \epsilon^{\mu\nu\rho\sigma}J^{\ast}_{\nu}J_{\rho}N_{\sigma}.
\end{equation}


The hadronic current $J_{\mu}$ in the particular final state $X$ depends on the momenta of all hadrons and the decay model of a mediated resonance.

In the case of $\tau \rightarrow \pi\nu$ the hadronic current is $J_{\mu} = f_{\pi}Q_{\mu}$ and the polarimetric vector~(\ref{polarimetricvector}) is

\begin{equation}\label{polarimetricpi}
\vec{h} = -\vec{n}_{\pi},
\end{equation}
where  $\vec{n}_{\pi}$ is a unit vector pointing along the direction of flight of the pion.

In the decay $\tau \rightarrow \rho\nu$  the  current $J_{\mu} = F_{\rho}q_{\mu}$ and the polarimetric vector~(\ref{polarimetricvector}) reads:

\begin{equation}\label{polarimetricrho}
\vec{h} = m_{\tau}\frac{2(qN)\vec{q} - q^{2}\vec{N}}{2(qN)(qP)  - q^{2}(NP)}.
\end{equation}

The situation in $\tau \rightarrow 3\pi\nu$ is somewhat more complicated due to the complexity of the resonance structure.
The explicit expression of the polarimetric vector in this decay is not given here, this can be calculated from~(\ref{polarimetricvector}) using
the corresponding hadronic current which is discussed in Sec.~\ref{decaymodel_syst}.

The angle $\cos\theta_{h}$  corresponds to the so-called an optimal observable~\cite{Davier:1992nw}. Consider a general form of the decay distribution with polarization $P_{\tau}$:

\begin{equation}\label{gammadistr}
\frac{1}{\Gamma}\frac{d\Gamma}{d\vec{\xi}} = f(\vec{\xi})  + P_{\tau}g(\vec{\xi}),
\end{equation}
with the normalization and positivity conditions: 
\begin{equation} \label{eq3}
\begin{split}
\int f(\vec{\xi}) d\vec{\xi}  = 1,\\
\int g(\vec{\xi}) d\vec{\xi}  = 0, \\
f(\vec{\xi}) \ge 0, \\
|g(\vec{\xi})| \le f(\vec{\xi}),
\end{split}
\end{equation}
where $\vec{\xi}$  is the vector of all spin sensitive angles.

\begin{figure}[h]
  \begin{center}
    \includegraphics[width=0.75\textwidth]{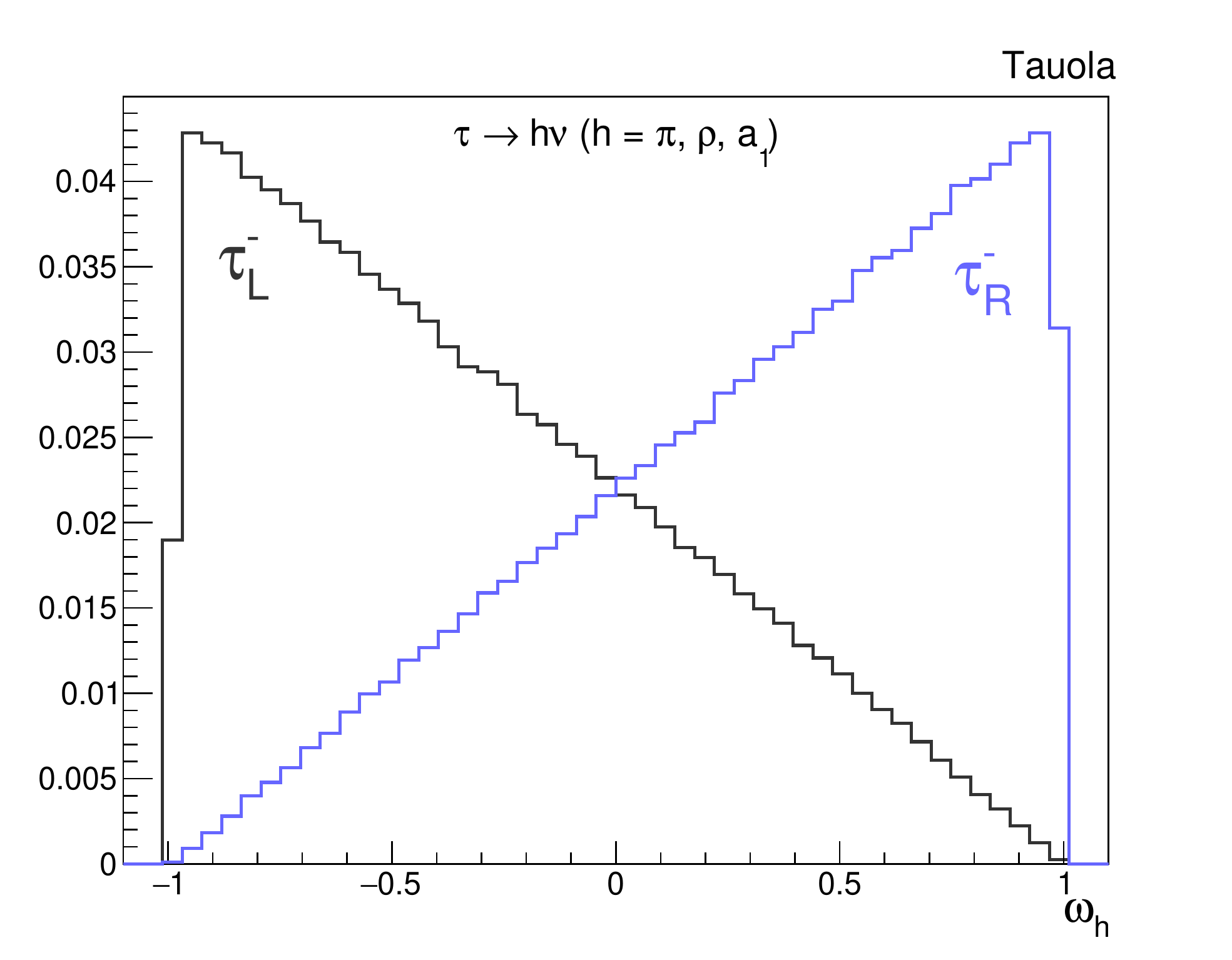}  
    \caption{The distribution of $\cos\theta_{h}$ for negative (black)  and positive (cyan) $\tau$ helicity. The distribution is identical for the $\tau$ decays to: $\tau \rightarrow \pi\nu$, $\tau \rightarrow \rho\nu$, $\tau \rightarrow a_{1}\nu$. }
      \label{omegah}
  \end{center}
\end{figure}

The general approach to extract the polarization in an optimal way  uses the advantage
of the linear dependence of the decay distributions on the polarization $P_{\tau}$. The true polarization maximizes 
the likelihood~(\ref{gammadistr}) $\mathcal{L}(P_{\tau},\vec{\xi})$ and therefore its value is obtained from:

\begin{equation}\label{loglike}
\begin{split}
\frac{d}{dP_{\tau}} log\mathcal{L}(P_{\tau},\vec{\xi}) = \sum\limits^{N}_{i}\frac{g(\vec{\xi_{i}})}{f(\vec{\xi_{i}}) + P_{\tau}g(\vec{\xi_{i}})} \\
=  \sum\limits^{N}_{i}\frac{\omega_{i}}{1 + P_{\tau}\omega_{i}} =0, 
\end{split}
\end{equation}
where $\omega = \frac{g(\vec{\xi})}{f(\vec{\xi})}$. If all decay products are reconstructed then for  any decay channel of $\tau$ from~(\ref{decaydistr2}) one obtains $f = \frac{1}{2}$ and $g =\frac{1}{2}\cos\theta_{h}$.
The distribution of $\cos\theta_h$ is shown in Fig.~\ref{omegah}. Events are generated  using Pythia8~\cite{Sjostrand:2014zea} supplemented with the TAUOLA~\cite{Jadach:1993hs} program for $\tau$ decays.

The gap in the first and last bins is caused by the finite mass of the hadron system.

The variance on $P_{\tau}$ is (in the limit of large $N$),

\begin{equation}
\frac{1}{\sigma^{2}} = -\frac{\partial^{2} log\mathcal{L}}{\partial^{2}P_{\tau}} = N\langle \frac{\omega^{2}}{(1+P_{0}\omega)^{2}} \rangle.
\end{equation}

As an assessment of the sensitivity of the given $\tau$ decay channel to the polarization the following parameter is defined, 

\begin{equation}\label{senstivitydefitions}
S^{2} = \frac{1}{N\sigma^{2}} = \langle \frac{\omega^{2}}{(1+P_{\tau}\omega)^{2}} \rangle = \int \frac{g^{2}}{f+P_{\tau}g}d\vec\xi.
\end{equation}

The parameter S quantifies the  statistical error that results from a likelihood fit to the ideal distributions, i.e. not altered by the reconstruction effects.
The sensitivity depends on the  $\tau$ decay channel analyzed and the  polarization  value $P_{\tau}$. As can be seen from~(\ref{senstivitydefitions})  a $\tau$ decay  channel with larger sensitivity 
provides a better statistical uncertainty on the extracted value of $P_{\tau}$.

In case of the full kinematic  of $\tau$ lepton is reconstructed the sensitivity reaches the maximum value  $\simeq $ 0.58 irrespective of the decay channel. The described approach provides an uniform description for the longitudinal $\tau$ spin analysis in any decay channel. 

\section{Spin correlations of $\tau^{+}\tau^{-}$ pair}\label{combined}

The helicities of two $\tau$ leptons in the reaction $Z \rightarrow \tau\tau$ are 100\% anti-correlated (in the limit $\frac{m_{\tau}}{E_{\tau}}$). 
Thus, the helicity states separation and consequently the sensitivity to the polarization measurement can be gain by exploiting this correlation. 
The method to take into account, proposed in Ref.~\cite{Davier:1992nw} consist in defining the new optimal observable for the whole event. Denoting $\omega_{1}$ 
and $\omega_{2}$ to be the observables for  $\tau^{+}$ and $\tau^{-}$, from the joint decay distribution one can obtain:

\begin{equation}
\Omega = \frac{\omega_{1} + \omega_{2}}{1 + \omega_1\omega_2}.
\end{equation}

\begin{figure}[ht]
  \begin{center}
    \includegraphics[width=0.75\textwidth]{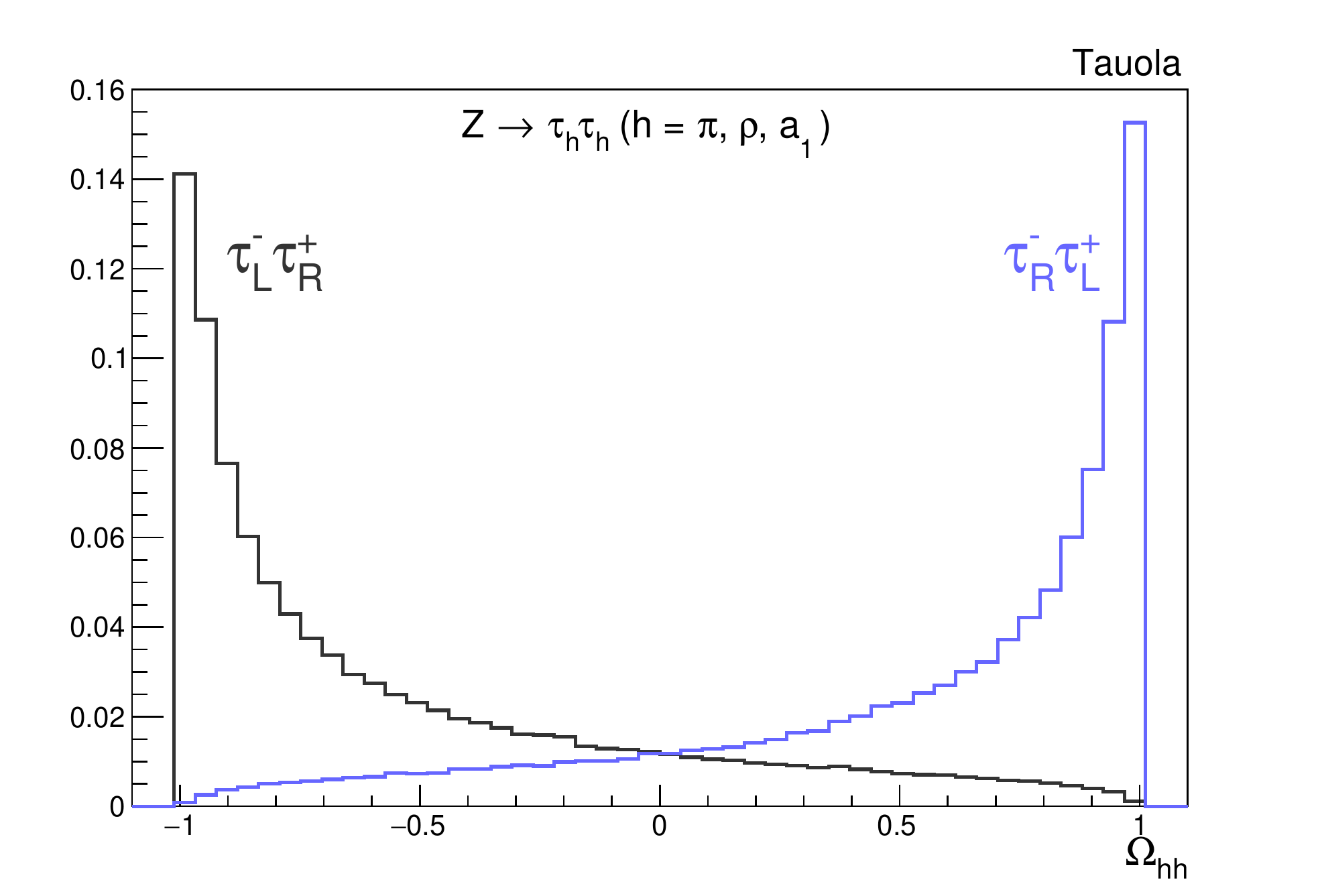}   
    \caption{The distribution of $\Omega_{hh}$ for negative (black)  and positive (cyan) $\tau$ helicity. The distribution is identical for any combination of hadronic decays $\tau_{h}\tau_{h}$ ($h=\pi$, $\rho$, $a_{1}$). }
    \label{Omegahh}
  \end{center}
\end{figure}

The distribution of $\Omega$ when both $\tau$ leptons decays into hadrons is shown in Fig.~\ref{Omegahh}. The separation of $\tau_{R}$ and $\tau_{L}$ is clearly improved comparing to distribution of $\omega_{h}$.

The sensitivity of this quantity reaches the value $\simeq $ 0.73, for any combination of hadronic decays of $\tau^{+}$ and $\tau^{-}$. This
can be compared to 0.58 in case of analyzing only one $\tau$ lepton.

The strategy of extracting $\tau$ polarization asymmetry from data consists in  performing a fit to the distribution of the variable $\omega_{h}$ in case 
of single $\tau$ analysis or a combined distribution $\Omega_{hh}$ in case of a $\tau$ pair analysis, measured from the data,
using two template distributions  for $\tau$ leptons with helicity $+1$ and $-1$ obtained from Monte Carlo simulation, with their relative fraction as a free parameter.

\section{$\tau \rightarrow 3\pi\nu$ decay model and theoretical uncertainty from the model dependence}\label{decaymodel_syst}

The $\tau$ decay to three pions and a neutrino is dominated by the $a_{1}(1260)$ ($J^{PG} = 1^{+-}$). It decays through the intermediate
state of ($\rho\pi$ ), with mostly $\rho(770)$ and an admixture of $\rho(1450)$ at higher
masses, followed by $\rho \rightarrow \pi\pi$ decay. One expects the two charge combinations
$\tau^{-} \rightarrow \pi^{+}\pi^{-}\pi^{-}\nu_{\tau}$ and $\tau^{-} \rightarrow \pi^{-}\pi^{0}\pi^{0}\nu_{\tau}$ to have a similar structure.

There are many experimental studies of $\tau^{-} \rightarrow \pi^{+}\pi^{-}\pi^{-}\nu_{\tau}$ were performed~\cite{Albrecht:1992ka,Ackerstaff:1997dv,Abreu:1998cn}
based on different theoretical approaches to describe  the decay $a_{1} \rightarrow 3\pi $~\cite{Kuhn:1992nz,Isgur:1988vm,Feindt:1990ev,Bowler:1988kf}. The best  satisfactory fits of the experimental spectra were presented by the CLEO collaboration~\cite{Asner:1999kj}. In total seven resonances were included in the fit to provide a good description if the CLEO data.  Table~\ref{tab:table1} shows the amplitudes and their contributions to the neutral mode.


\begin{table}[h!]
  \begin{center}
    \caption{Result of a fit of the $a_1$ including different intermediate resonances to the decay $\tau^{-}\rightarrow \pi^{-}\pi^{0}\pi^{0}\nu_{\tau}$ be the CLEO collaboration~\cite{Asner:1999kj}. The first column gives the intermediate state, the second specifies the relative angular momentum between two particles, the third is the significance of the amplitude in the standard deviations and the last column is the fraction of the events.}
    \label{tab:table1}
    \begin{tabular}{l|ccc} 
      {Resonance} & {L} & {Significance} & {Fraction, \%}\\
      \hline
      $\rho(770)\pi$ & S  &  -  & 68.11\\
      $\rho(1450)\pi$ & S & 1.4$\sigma$  & 0.30 $\pm$ 0.66\\
      $\rho(770)\pi$ & D & 5.0$\sigma$  & 0.36 $\pm$ 0.18\\
      $\rho(1450)\pi$ & D  & 3.1$\sigma$  & 0.43 $\pm$ 0.29\\
      $f_{2}(1270)\pi$ & P & 4.2$\sigma$ & 0.14 $\pm$ 0.06\\
      $\sigma\pi$ & P & 8.2$\sigma$  & 16.18 $\pm$ 4.06\\
      $f_{0}(1370)\pi$ & P & 5.4$\sigma$ & 4.29 $\pm$ 2.40\\

    \end{tabular}
  \end{center}
\end{table}

In order to calculate the polarimetric vector for the decay $\tau \rightarrow (3\pi)\nu$ one needs to consider all these seven resonances.  The hadronic current in the decay $\tau \rightarrow 3\pi\nu$ including all intermediate contributions reads as:

\begin{equation}\label{hadrcurra1}
J_{a_{1}}^{\mu} = B_{a_1}(s)\times\sum_{i}\beta_{i}j_{i}^{\mu},
\end{equation}
where $B_{a_1}(s)$ denotes the $a_{1}$ Breight-Wigner, $\beta_{i}$ are complex coupling constants, and $j_{i}^{\mu}$ are formfactors describing the substructure involving the particular resonance.
An explicit form of Breight-Wigner functions and details of this parametrization can be found in Ref.~\cite{Asner:1999kj}.  This should be noted that the  model  parametrization of the $\tau \rightarrow 3\pi\nu$ decay (resonance parameters, couplings and/or intermediate resonances) in various Monte Carlo generators can be different from those described in CLEO paper. Therefore for a correct calculation of the polarimetric vector one should explicitly use the model introduced in particular MC generator intended to be used for a data analysis. A comprehensive review of the variants for Monte Carlo generators of the $\tau \rightarrow 3\pi\nu$ decay can be found in Ref.~\cite{Was:2015laa}.

As it was mentioned above the polarimetric vector in the decay $\tau \rightarrow 3\pi\nu$  strongly depends on the decay model of $a_{1}$. 
Thus, this is important to understand possible biases on the $\tau$ polarization measurement coming from the model dependence of the hadronic current. 
The lineshape of $a_{1}$ resonance enters similarly into the nominator and denominator of the Eq.~(\ref{polarimetricvector})  and therefore brings no impact. The theoretical systematic uncertainty comes from the imperfect knowledge of the substructure of the decay $a_{1} \rightarrow 3\pi$. The procedure used to estimate the uncertainty is the following.  Distributions
of the $\cos\theta_{h}$ for a positive and negative $\tau$ helicity have been produced with the TAUOLA Monte Carlo program using the nominal values for the  parameters and  relative contribution
of the resonances listed in Table~\ref{tab:table1}.  Further these distributions will be referred to as  $\cos\theta^{R}_{h}$ and $\cos\theta^{L}_{h}$.  Then,  a linear combination of $\cos\theta_{h}^{R}$ and $\cos\theta_{h}^{L}$ is produced with a fixed value of the $\tau$ polarization ($P^{0}_{\tau} = -0.15$).  Two additional samples of $\tau \rightarrow 3\pi\nu_{\tau}$ decay have been 
generated with a modified version of TAUOLA corresponding to "up" and "down" variation of the resonance parameters and their relative contribution. Parameters were varied within their fit uncertainties given by CLEO. Two fits of the templates for  $\cos\theta_{h}^{R}$ and $\cos\theta_{h}^{L}$  taken from the samples produced  with varied parameters to the sample with nominal parameters values have been performed. The half difference of the value for $P^{0}_{\tau}$ extracted from both fits  is taken as an estimate of the uncertainty.  The obtained variation is  $(\Delta P_{\tau})_{model} = (1.41 \pm 1.37)\times 10^{-4}$.  The uncertainty on this value comes from the Monte Carlo statistics used to perform the fits.

\section{Determination of the effective weak mixing angle}\label{determineSinEff}

The tree-level relation between $\tau$ polarization and the effective electroweak mixing angle $\cos\theta_{eff}$~(\ref{eqnumber5}) holds only at $Z^{0}$ - pole. 
However, in proton-proton collisions the center-of-mass energy of initial quarks in the process $q\bar{q} \rightarrow Z^{0}$ is not know. In this section a method 
to propagate the measured $\tau$ polarization in proton-proton collision including radiative corrections is discussed.

The  $\tau$ polarization $P_{\tau}(s)$ and the cross section $\sigma(s)$  can be expressed in terms of the structure functions from  Eq.~(\ref{crossection}):

\begin{equation}\label{eq:polxsecandstruc}
\begin{split}
\sigma_{q} = \frac{16}{3}F^{q}_{0}(s), \\
P^{q}_{\tau}(s) =  -\frac{F^{q}_{2}(s)}{F^{q}_{0}(s)},
\end{split}
\end{equation}
where label $q$ denotes the flavour of the initial quark anti-quark pair in the process $q\bar{q} \rightarrow Z \rightarrow \tau\tau$. 
Structure function $F_{i}^{q}$ contain the electrical charge and weak coupling constants of quarks and therefore away from the $Z^{0}$ pole the $\tau$ polarization depends on the initial quark flavour. 
 
\begin{figure}[t]
  \begin{center}
    \includegraphics[width=0.650\textwidth]{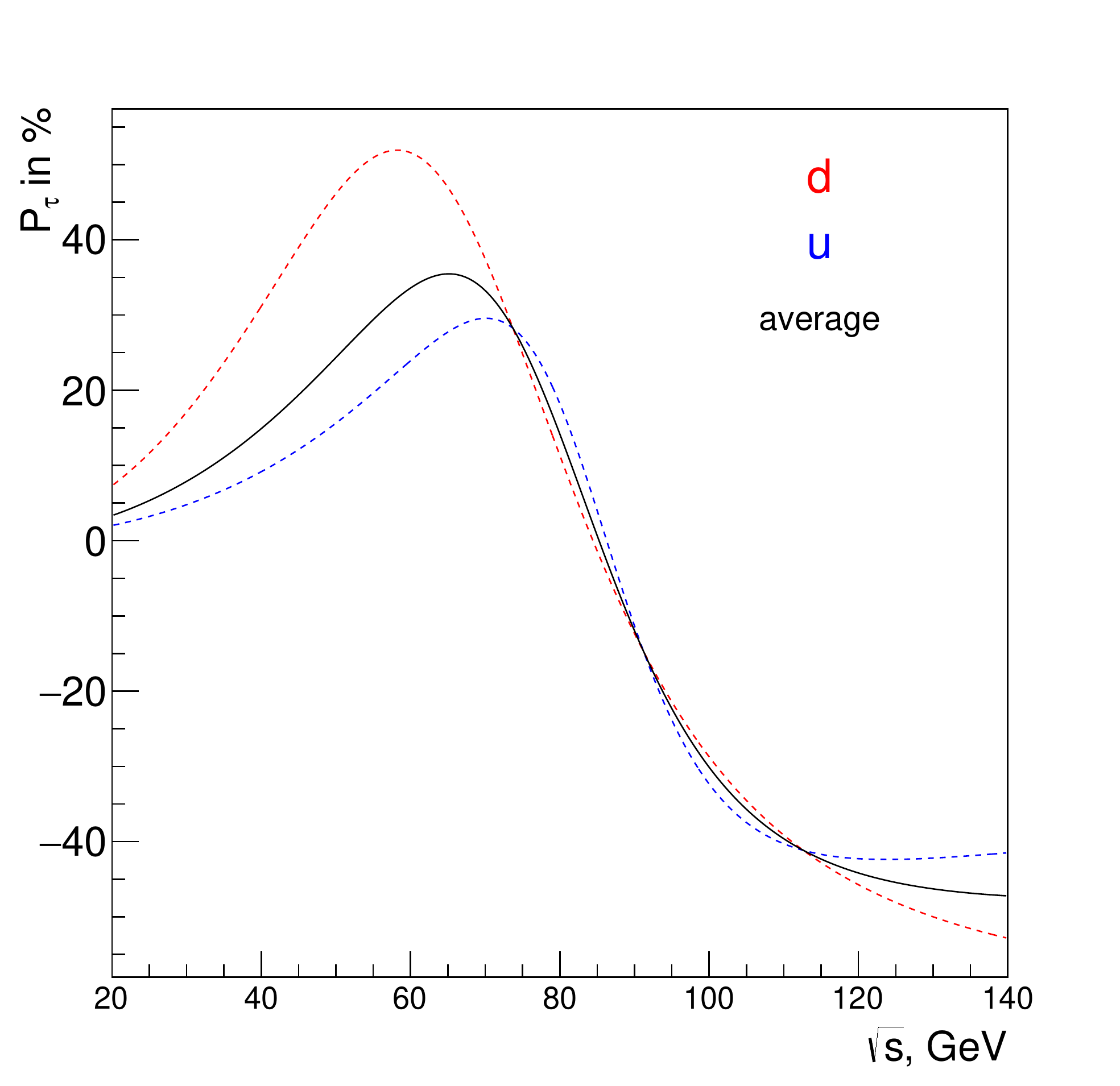}
  \caption{The $\tau$ polarization as a function of $\sqrt{s}$. \textit{Dashed blue curve} polarization for up-type quarks (u, c) in the initial state, \textit{Dashed red curve} for down-type quarks (d, s, b). \textit{Solid curve} the average with the factor $\alpha_{u} = 0.423$ taken from Madgraph Monte Carlo program~\cite{Alwall:2014hca}. The parameters  $m_{Z} = 91.1867 GeV$, $\Gamma_{Z} = 2.4939 GeV$, $\sin^{2}\theta_{W} = 0.23155$ are used in the tree-level calculation.       }
    \label{fig:ew:tauProd1}
  \end{center}
\end{figure}

The $\tau$ polarization as a function of the center-of-mass energy of the quark - antiquark pair $\sqrt{s}$, for up-type quarks (u ,c), down-type (d, s, b) and their average shown in Fig.~\ref{fig:ew:tauProd1}.  The $\tau$ polarizations for u- and d- quark types are crossing each other at the point $\sqrt{s} = M_{Z}$ since at the Z pole the polarization does not depend on the initial flavour.
The average polarization is calculation using  the factor $\alpha$  that  describes the relative contribution of up and down type quarks in the  Z formation. The numerical value  for $\alpha = 0.423$
is taken from the Madgraph Monte Carlo program~\cite{Alwall:2014hca}.

Denoting $\hat{P}^{u}_{\tau}$ the $\tau$ polarization in the process where the $Z$ boson is formed in annihilation of up-type quarks (u, c), and the $\hat{P}^{d}_{\tau}$ to be the polarization from
annihilation of down-type quarks (d, s, b) the $\tau$ polarization measured in proton-proton collisions averaged  over the center-of-mass energy of the $\tau$ pair and the the initial quarks flavour is:

\begin{equation}\label{averpolfin}
\begin{split}
&<P_{\tau}>(\cos\theta_{eff}) = \\
&\frac{\int [\alpha\hat{P}^{u}_{\tau}(x)\sum\limits_{q=u, c}\hat{\sigma}_{q}(x) + (1-\alpha)\hat{P}^{d}_{\tau}(x)\sum\limits_{q=d, s, b}\hat{\sigma}_{q}(x)]\epsilon(x)dx}{\int\sum\limits_{q=u, c, d, s, b}\hat{\sigma}_{q}(x)\epsilon(x)dx},\\
\end{split}
\end{equation}
where $x = \sqrt{s}$,    $\epsilon(x)$  is the acceptance, $\hat{\sigma}_{q}(x)$ is the cross section of the process $pp \rightarrow Z^{0} \rightarrow \tau^{+}\tau^{-}$ for a given initial quarks flavour $q$.
The cross section $\hat{\sigma}_{q}(s)$  is related to the cross section $\sigma_{q}(s)$ of the subprocess $q\bar{q} \rightarrow Z^{0} \rightarrow \tau^{+}\tau^{-}$ by weighting the last
with  the parton distribution functions (PDF) $f_{q}(s)$ and $f_{\bar{q}}(s)$. This relation in the Leading Order (LO) by $\alpha_{s}$ reads~\cite{Drell:1970yt}:

\begin{equation}\label{NLODRELLYANXSEC}
\begin{split}
&\frac{d\hat{\sigma}_{q}^{LO}(s)}{ds} =\int\limits_0^1 dx_{1}dx_{2} \left \{ f_{q}(x_{1},s) f_{\bar{q}}(x_{2},s) +  (q \leftrightarrow \bar{q})  \right \}\frac{d\sigma_{q}(s)}{ds}.  \\
\end{split}
\end{equation}

 The PDFs describe the probability density distribution of partons (quarks, antiquarks or gluons) as a function of $x_1$ and $x_2$, the momentum fractions of a hadron carried by a parton.

 For a correct calculation of the integral in Eq.~(\ref{averpolfin}) this is important to evaluate next-to-leading (NLO) order corrections to the cross section $\hat{\sigma_{q}}(s)$.
 The next-to-leading QCD orders  can be written as perturbative expansion of $\mathcal{F}$ in powers of the strong coupling $\alpha_{S}$:
\begin{equation}\label{LOtoNLO}
\begin{split}
&\frac{d\hat{\sigma}^{NLO}_{q}(s)}{ds}  = \frac{\hat{\sigma}_{q}^{LO}(s)}{s}\kappa\mathcal{F}(\kappa),\\
&\mathcal{F}(\kappa) = \mathcal{F}_{0}(\kappa) + \frac{\alpha_{S}}{2\pi}\mathcal{F}_{1}(\kappa) + \ldots,\\
\end{split}
\end{equation}
where  $\kappa = s/S$, $S$ is the square of the hadron-hadron collision energy. An explicit form of $\mathcal{F}_{1}(\kappa)$ was studied by many authors and can be found for example
in Ref.~\cite{ellis2003qcd}.

\begin{figure}[t!]
  \begin{center}
    \includegraphics[width=0.650\textwidth]{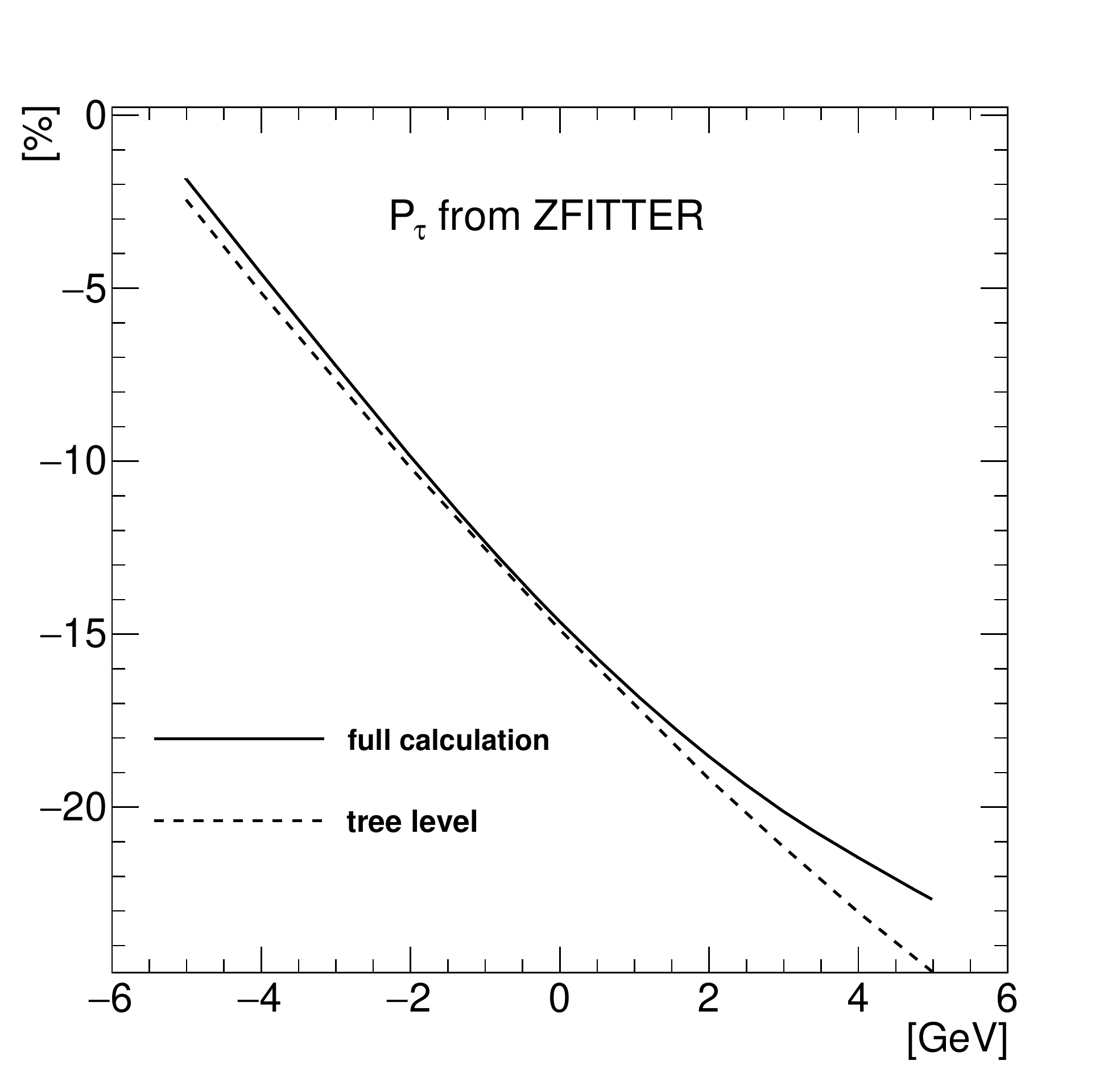}
  \caption{$\tau$ polarization as a function of center of mass energy relative to the $Z^{0}$ mass. \textit{Dashed line}: tree level calculation; \textit{Solid line}: full radiative corrections for the process $q\bar{q} \rightarrow Z^{0} \rightarrow \tau\tau$ obtained from ZFitter.}
    \label{RadCorrToTauPol}
  \end{center}
\end{figure}

The electroweak radiative corrections for $\sigma_{q}$ and $P^{u,d}_{\tau}$ can be calculated using ZFitter program~\cite{Akhundov:2013ons}.
ZFitter  package was developed to calculate the radiative corrections, as predicted in the Standard Model, 
to a variety of observable quantities, related to the Z-boson resonance produced in $e^{+}e^{-}$ collisions, however by 
changing the relevant parameters (charge and weak couplings of the particles in the initial state) it can modified to calculate the  quantities of interest for the Z boson produced in the annihilation of $q\bar{q}$. 
The effect of radiative correction on the $\tau$ polarization calculated using ZFitter is shown in Fig.~\ref{RadCorrToTauPol}.

The method that we propose here to determine the effective weak mixing angle from  the measured in data average $\tau$  polarization $<P_{\tau}>$ 
consists in evaluation of the integral~(\ref{averpolfin}) with values for $\sigma_{q}$ and $P^{u,d}_{\tau}$  that correspond to a range of values for $\sin^{2}\theta_{eff}$. 
The value for $\sin^{2}\theta_{eff}$ can be set manually in ZFitter, for each value one can calculate the expected average $\tau$  polarization $<P_{\tau}>$ taking into account
the uncertainties on PDFs,  the theoretical uncertainty on the EWK parameters calculated by ZFitter and the uncertainty on the acceptance $\epsilon(s)$.  The acceptance efficiency
$\epsilon(s)$  specifically depends on the selections cuts applied to data sample and can be studied using corresponding  Monte Carlo sample of Drell-Yan events. The uncertainty on
$\epsilon(s)$ comes from the limitation on Monte Carlo statistics used for the analysis.  

\begin{figure}[hbtp]
  \begin{center}
    \includegraphics[width=0.60\textwidth]{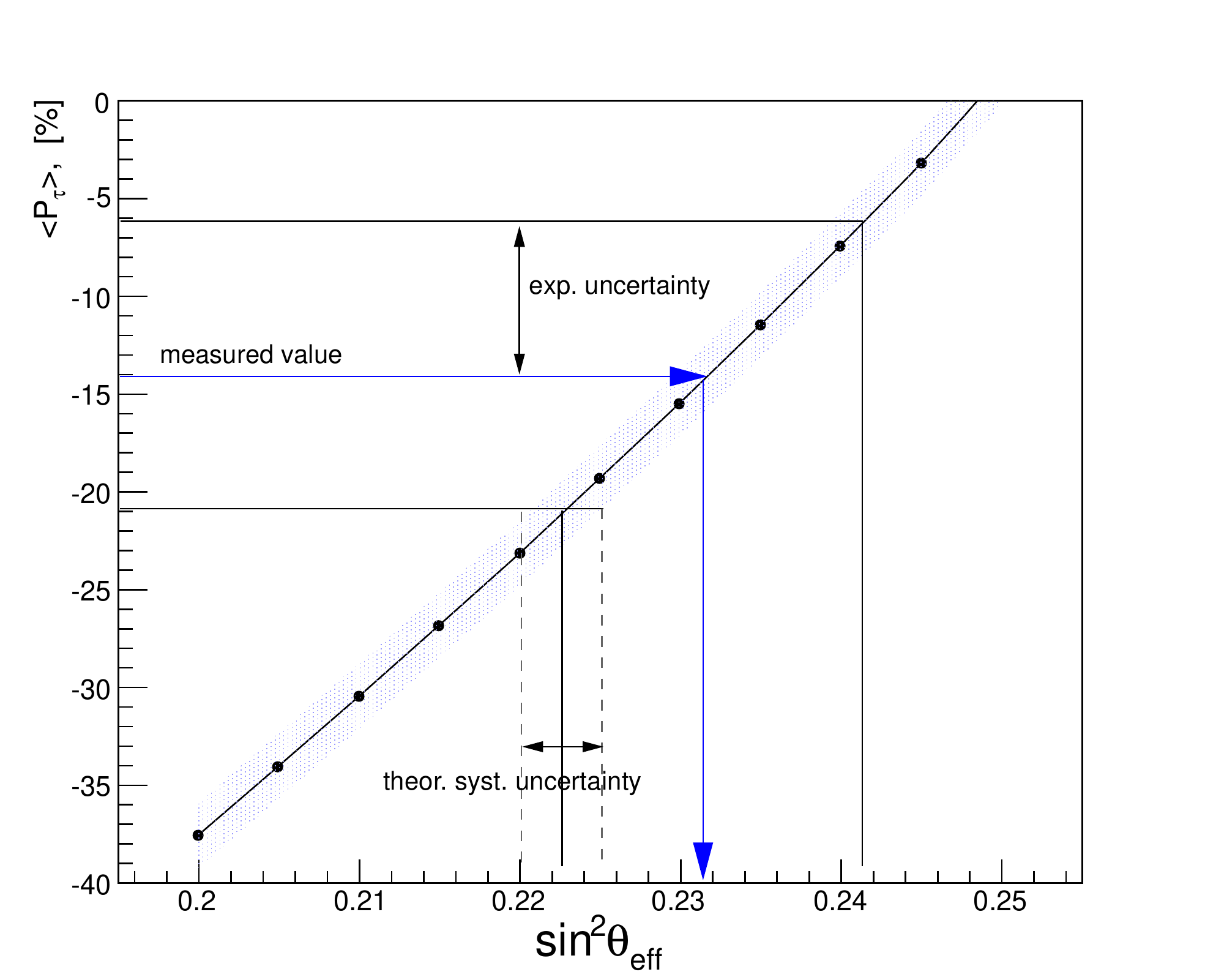}
   
    \caption{The average $\tau$ polarization as a function of the weak mixing angle. The dashed band represent the theoretical uncertainty that includes uncertainties of PDFs, theoretical uncertainties of Drell-Yan cross section and theoretical uncertainty on $\tau$ polarization. }
    \label{avervsange}
  \end{center}
\end{figure}

The value of the measured average $\tau$ polarization as well as its uncertainty then can be propagated to the value of the effective mixing angle, as schematically shown in Fig.~\ref{avervsange}.

The ratio of effective vector and axial-vector couplings of $\tau$  leptons can be determined then from the relation:

\begin{equation}\label{effcouplandangle}
\frac{\bar{v}_{\tau}}{\bar{a}_{\tau}} = 1 -4|Q_{\tau}|\sin^{2}\theta_{eff},
\end{equation}
where $\bar{v}_{\tau}$ and $\bar{a}_{\tau}$ are the effective weak vector and axial-vector couplings of $\tau$ lepton.

This should be noted that numerically the difference between the descibed propagation and the tree level is not large comparing to the possible precision of $\tau$ polarization measurement that can be recently reached by
 CMS or ATLAS detectors.
This can be seen in Fig.~\ref{fig:ew:tauProd1} the $\tau$ polarization below the $Z$ pole and above will have an opposite sign in the integration~(\ref{averpolfin}) and thus partially cancel each other. 
Therefore, the tree level propagation~(\ref{eqnumber5}) can be used to a very good approxiamtion.

\section{Conclusions}

A general method to measure $\tau$ lepton  polarization in the process $pp \rightarrow Z^{0} \rightarrow \tau\tau$ is described. Assuming the
knowledge of the full kinematic of the $\tau$ pair, which can be estimated by various methods, the statistical uncertainty 
of this measurement can be maximized by reconstructing the optimal observable for each decay. The sensitivity and as a consequence the statistical uncertainty of the extracted value for
$\tau$ polarization can further be gained exploiting the longitudinal spin correaltion of $\tau$ pair. An estimate of the systematic uncertainty $(\Delta P_{\tau})_{model} = (1.41 \pm 1.37)\times 10^{-4}$ due to the model 
dependence in the $\tau \rightarrow a_{1}\nu \rightarrow 3\pi\nu$ decay has been obtained. This uncertainty is far below the current precision of the $\tau$ polarization and therefore will not limit the precision in this channel.
The effective electroweak mixing angle  can  be determined from  the measured average $\tau$ polarization taking into account the EWK radiative effects, parton density functions and NLO QCD corrections to the Drell-Yan cross section.

\section{Acknowledgments}

The authors are very grateful to Zbigniew Was, Tord Riemann and Mikhail Pletyukhov for very fruitful  and useful conversations. 
We would like to acknowledge Thomas M{\"u}ller and Johannes Merz for  comprehensive checks of the presented results.  We also wish to thank Yannick Patois for computing support.


\end{document}